\documentstyle[12pt]{article}
\input epsf
\begin{document}
\rightline{FERMILAB-PUB-94/298-T}
\rightline{EFI-94-43}
\rightline{October 1994}
\rightline{hep-ph/9411213}
\vspace{0.7in}
\centerline{\bf ISOSPIN CONSIDERATIONS IN CORRELATIONS OF PIONS}
\centerline{\bf AND $B$ MESONS}
\vspace{0.5in}
\centerline{\it Isard Dunietz}
\centerline{\it Theory Group, Fermilab}
\centerline{\it P. O. Box 500, Batavia, IL 60510}
\medskip
\centerline{and}
\medskip
\centerline{\it Jonathan L. Rosner}
\centerline{\it Enrico Fermi Institute and Department of Physics}
\centerline{\it University of Chicago, Chicago, IL 60637}
\vspace{0.5in}
\centerline{\bf ABSTRACT}
\medskip
\begin{quote}
The correlations between a $B$ meson and a pion produced nearby in phase space
should respect isospin reflection symmetry $I_3 \to -I_3$.  Thus, one generally
expects similar $\pi^+ B^0$ and $\pi^- B^+$ correlations (non-exotic channels),
and similar $\pi^- B^0$ and $\pi^+ B^+$ correlations (exotic channels).
Exceptions include (a) fragmentation processes involving exchange of quarks
with the producing system, (b) misidentification of charged kaons as charged
pions, and (c) effects of decay products of the associated $\overline{B}$. All
of these can affect the apparent signal for correlations of charged $B$ mesons
with charged hadrons.  The identification of the flavor of neutral $B$
mesons through the decay $B^0 \to K^{*0} J/\psi$ requires good particle
identification in order that the decay $K^{*0} \to K^+ \pi^-$ not be mistaken
for $\overline{K}^{*0} \to K^- \pi^+$, in which case the correlations of
neutral $B$ mesons with hadrons can be underestimated.
\end{quote}
\newpage

\centerline{\bf I.  INTRODUCTION}
\bigskip

It has recently been suggested \cite{GNR} that one can identify the flavor of a
neutral $B$ meson at the time of production through its correlation with a
charged pion produced nearby in phase space.  If such correlations are found,
one has a useful means of studying CP-violating rate asymmetries in the decays
of neutral $B$ mesons to CP eigenstates like $J/\psi K_S$.

In order to calibrate the effectiveness of this method, one can study the
correlations between charged pions and $B$ mesons of {\it known} flavor. Most
decay modes of neutral $B$'s, such as $B^0 \to D^{*-} \ell^+ \nu_\ell$ and $B^0
\to K^{*0} J/\psi$, provide flavor information.  However, the reconstruction of
a $B$ is hindered by the absence of the neutrino in the first case, while in
the second there is the potential of confusing the decay $K^{*0} \to K^+ \pi^-$
with $\overline{K}^{*0} \to K^- \pi^+$ if particle identification is not
efficient.

The flavor of a charged $B$ is inferred directly from its decay mode, such as
$J/\psi K^{\pm}$.  Thus, one can study the correlations between pions and
charged $B$ mesons with relative ease, and it was argued in Refs.~\cite{GNR}
that such correlations should shed light on the corresponding correlations
involving neutral $B$ mesons.  In this article we wish to explore the relations
between charged- and neutral-$B$ correlations with pions more fully.

One might have expected the correlations involving charged and neutral $B$
mesons to be different just by considering the overall charge. For example, in
reactions with overall charge zero such as $e^+ e^-$ or $\bar p p$ collisions,
the particles accompanying a $B^+$ have total charge $-1$, while those
accompanying a $B^0$ have total charge zero.  Thus, there ought to be more
negative pions available for correlation with a $B^+$ than there are positive
pions available for correlation with a $B^0$.  Nonetheless~\cite{GNR}, we
expect under normal circumstances the correlations in the non-exotic channels
$\pi^+ B^0$ and $\pi^- B^+$ to be identical, and stronger than those in the
exotic ones $\pi^- B^0$ and $\pi^+ B^+$, which should also be identical
\cite{EX}.  This argument, reviewed in Sec.~II, is a simple consequence of
invariance under isospin reflection, or equivalently of the interchange of
nonstrange quarks $u \leftrightarrow d$.  It should be valid as long as the $B$
mesons are produced by fragmentation from an initial $b \bar b$ system which
does not exchange flavor quantum numbers with the rest of the production
process.

Exchange of quarks with the rest of the system which gives rise to the $b \bar
b$ pair could invalidate the argument, as discussed in Sec.~III\null. Several
other instrumental effects can give an {\it apparent} difference between
charged- and neutral-$B$ correlations.  These include the misidentification of
charged kaons as charged pions (Sec.~IV),  the effects of pions in the decay of
the associated $\overline{B}$ meson (Sec.~V), and the misidentification of the
flavor of a neutral $B$ as a result of inadequate particle identification
(Sec.~VI).  We show how to test for each of these effects.  Sec.~VII concludes.
\newpage

\centerline{\bf II.  EQUALITY OF CORRELATIONS WITH PIONS}
\bigskip

We consider here only associated production of $B$ hadrons, arising as a result
of fragmentation of an initial $b \bar b$ pair.  In general this pair will
arise either from one or more gluons (as in hadronic production), from a
photon and one or more gluons (as in photoproduction), from the
interference of a virtual photon and a virtual $Z$ as in electron-positron
collisions or a Drell-Yan process, or through the decay of a real $Z$.

As long as the $b \bar b$ pair is produced in isolation from other hadrons
containing light quarks, the fragmentation of a $b$ quark (Fig.~1) should
produce equal ``right-sign'' correlations between charged pions and $B$ mesons
of either flavor.  Thus, if the $\bar b$ quark in Fig.~1 fragments into a $B^0$
by producing a quark $q = d$, the next hadron down the chain will contain a
$\bar d$.  If this hadron is a charged pion, it must be a $\pi^+$. If the $\bar
b$ fragments into a $B^+$ by producing a $u$ quark, and the next hadron down
the chain is a charged pion, it must be a $\pi^-$.  The two processes are
related to one another by the interchange $d \leftrightarrow u$ and clearly
have equal probabilities.

\begin{figure}
\centerline{\epsfysize = 2 in \epsffile {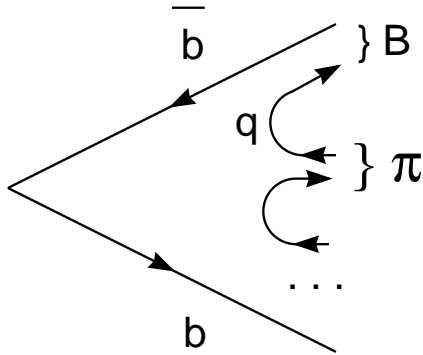}}
\caption{Fragmentation of a $\bar b b$ pair into a $B$ meson and other
particles in the absence of interaction with quarks in the rest of the system.}
\end{figure}

A similar conclusion can be drawn regarding the pions arising from decays of
excited ``$B^{**}$'' resonances.  As long as charged and neutral resonances are
produced with equal numbers, invariance under reflection of the third component
of isospin, $I_3$, requires their decays to give equal $\pi^+ B^0$ and $\pi^-
B^+$ correlations.  The exotic correlations $\pi^- B^0$ and $\pi^+ B^+$ also
should be equal.

The decay $Z \to b \bar b$ is an example of a process in which the above
conditions are expected to hold, with the pair fragmenting in an isospin
invariant manner.  A hadronically produced $b \bar b$ pair might also fragment
in an isospin-invariant manner as long as it does not interact with the rest of
the system.  For example, if two gluons collide to form a color-singlet $b \bar
b$ pair, a diagram very similar to Fig.~1 might be expected to account for the
hadronization of the heavy quarks.  (One should note that a pair of heavy
quarks produced in a hadronic collision is produced in a color-octet state a
significant amount of the time.)
\bigskip

\centerline{\bf III.  INTERACTION WITH THE PRODUCING SYSTEM}
\bigskip

A $b \bar b$ pair produced hadronically in a color-non-singlet state will
exchange color with the rest of the system upon hadronization.  Normally this
process will proceed in an isospin-invariant manner, via exchange of
one or more gluons or isosinglet quark-antiquark pairs.  However, there can be
exceptions to this rule, an example of which is shown in
Fig.~2.  A proton diffractively dissociates into a $b$-flavored baryon and a
meson containing a $\bar b$ quark.  Since the proton has more $u$ quarks than
$d$ quarks, the production of charged $B^{**}$ states is enhanced over that of
neutral ones.  In this example, $B^{**+}$ decays 2/3 of the time to $\pi^+
B^{(*)0}$ and 1/3 of the time to $\pi^0 B^{(*)+}$, while $B^{**0}$ decays 2/3
of the time to $\pi^- B^{(*)+}$ and 1/3 of the time to $\pi^0 B^{(*)0}$.

\begin{figure}
\centerline{\epsfysize = 1 in \epsffile {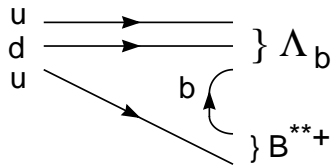}}
\caption{Diffractive dissociation of a proton into a $b$-flavored meson and
baryon, illustrating one source of unequal correlations involving charged and
neutral $B$ mesons with charged pions.}
\end{figure}

One can construct further examples in which an initial $b \bar b$ pair produced
by one or more gluons picks up quarks from the surrounding medium, as shown in
Fig.~3. Since the proton-antiproton system has states of both $I=0$ and $I=1$,
the rates for production of charged and neutral excited $B$ mesons can differ,
and so can the correlations between pions and charged and neutral $B$'s. One
might expect such effects, as well as the diffractive process in Fig.~2, to be
least important for central $B$ production at high momentum transfers, and most
important for processes in which the heavy quarks are produced at large
longitudinal and small transverse momenta.

\begin{figure}
\centerline{\epsfysize = 4 in \epsffile {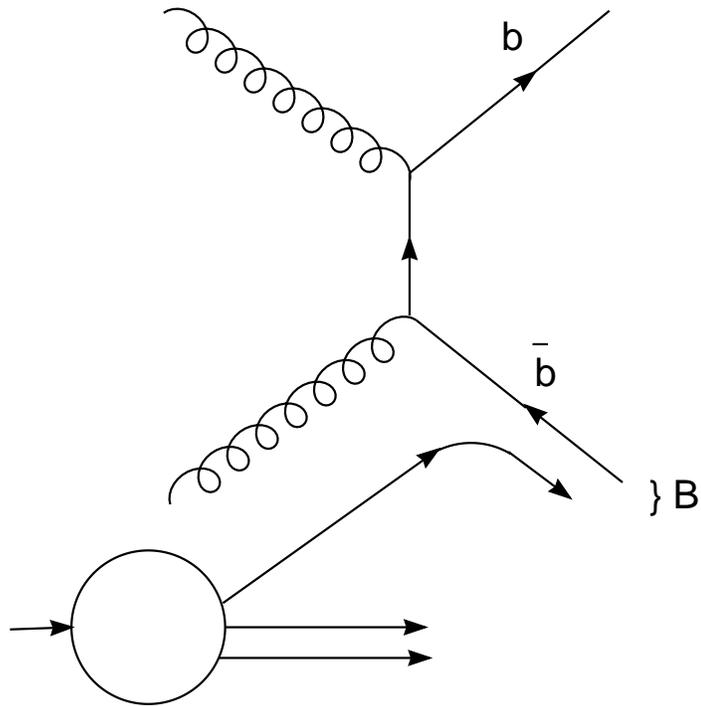}}
\caption{Production of a $b \bar b$ pair followed by exchange of quarks with
the producing system.}
\end{figure}

One expects processes such as those illustrated in Figs.~2 and 3 to be
important mainly for hadronic collisions.  In order that they contribute in
$e^+ e^-$ collisions, the light quarks must be produced directly by the
electroweak current.  For example, a $Z$ decays to $u \bar u$ and $d \bar d$
pairs with different rates.  If a $b \bar b$ pair is then produced by gluon
radiation, as shown in Fig.~4, there can be apparent violations of isospin
reflection symmetry in the fragmentation process.
\bigskip

\begin{figure}
\centerline{\epsfysize = 2 in \epsffile {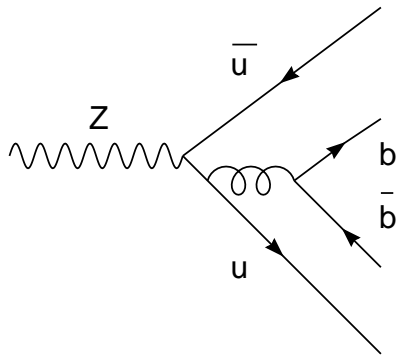}}
\caption{Decay of a $Z$ to $u \bar u$ followed by gluonic emission of a $b \bar
b$ pair.}
\end{figure}

\centerline{\bf IV.  EFFECTS OF ASSOCIATED CHARGED KAONS}
\bigskip

The correlations between charged $B$'s and charged hadrons nearby in phase
space can receive an important contribution from kaons, as shown in Fig.~5. No
such correlation with charged kaons exists for neutral $B$'s.  In the limit of
flavor SU(3) symmetry, in fact, the correlation between a charged $B$ and an
oppositely charged kaon would be {\it equal} to that between a charged $B$ and
an oppositely charged pion, effectively {\it doubling} the signal in comparison
with the $B^0 \pi^+$ or $\overline{B}^0 \pi^-$ correlation.

\begin{figure}
\centerline{\epsfysize = 2 in \epsffile {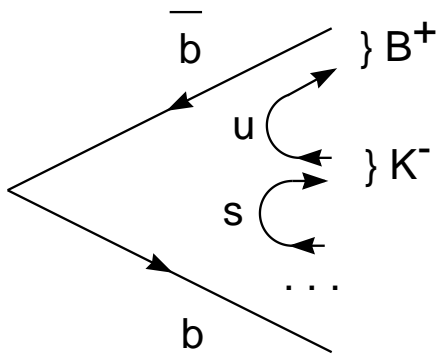}}
\caption{Fragmentation of a $\bar b$ quark into a $B^+$ meson and a $K^-$.}
\end{figure}

One can invent other processes in which the particle correlated with the $B$ is
a baryon.  Instead of the $s$ quark in Fig.~5, one would have an anti-diquark.
These examples illustrate the importance of particle identification over a wide
range of kinematic configurations.  If one can exclude processes such as shown
in Fig.~5, the correlation of charged $B$'s with charged hadrons becomes a much
more reliable tool for estimating similar correlations involving neutral $B$'s.
\bigskip

\centerline{\bf V.  EFFECTS OF PIONS IN DECAY OF ASSOCIATED $\overline{B}$}
\bigskip

The fragmentation ``chains'' as depicted in Figs.~1 and 5 have very different
lengths in high-energy $e^+ e^-$ and hadronic collisions.  In the decay of a
$Z^0$ to $\bar b b$, the heavy quarks and the products of their fragmentation
form two distinct jets whose components are highly unlikely to be confused with
one another.  However, in hadronic collisons, the $\bar b$ and $b$ are produced
with a spectrum of effective masses which peaks not far above threshold.  Thus,
unless one takes special care to ensure against it, the decay products of the
associated $\overline{B}$ can be among the pions which are correlated with the
$B$ of interest.

There are several sources of isospin violation in $B$ decays.  (a) The
transition $b \to c \bar u d$ changes $I_3$ by $-1$.  (b)  The transition $c
\to s u \bar d$ changes $I_3$ by $+1$.  (c)  The decays of charged and neutral
$D^*$'s violate isospin, since the channel $D^{*0} \to D^+ \pi^-$ is
kinematically forbidden. The effects of (a) and (b) largely cancel one another,
but (c) can play a significant role in skewing the expected charge
distributions of pions in $B$ decays.

The inclusive yield of charged pions in decays of $B$ and $\bar B$ mesons has
been measured for their sum [as produced at the $\Upsilon(4S)$], but not yet
individually \cite{BHP}, and not yet for baryons or antibaryons.
The importance of such a measurement is very great,
since it allows one to estimate the contamination of correlation signals by $b$
decay products.  One could guard against this effect if one could identify the
decay vertex of the second $b$. The pions of interest for performing
correlation studies come from the primary vertex and not from detached
vertices.
\bigskip

\centerline{\bf VI.  IDENTIFICATION OF NEUTRAL $K^*$ FLAVOR}
\bigskip

Neutral $B$'s of identified flavor have been fully reconstructed in many
different decay modes \cite{BHP}. However, the only easily accessible mode in
hadron colliders up to now has been $B^0 \to J/\psi K^{*0}$ because of
the ease of identification of the $J/\psi$.  One then identifies the
flavor of the $K^{*0}$ through its decay to $K^+ \pi^-$.

If particle identification is not efficient, a $K^+ \pi^-$ system
can be confused with $K^- \pi^+$, especially if only a fraction of the
available phase space is sampled.  To see this, let $p_\pi$  and $p_K$ be the
4-momenta of the pion and kaon, and let them have a squared invariant mass
$m_{K^*}^2 = (p_\pi + p_K)^2$.  If the pion and kaon are interchanged, the
error in the squared mass is approximately
\begin{equation}
\Delta m^2 \equiv m_{\rm incorrect}^2 - m_{\rm correct}^2
\simeq (m_K^2 - m_\pi^2)(\vec{p}_K^{~2} - \vec{p}_\pi^{~2})/
(|\vec{p}_\pi||\vec{p}_K|)~~~.
\end{equation}
Thus, one requires very asymmetric decay configurations in order to see a
broadening of a $K^*$ peak as a result of the wrong assignment of a pion and
kaon.  If any circumstances limit the momenta of pions and kaons accepted in
the data sample to a narrow range, the potential for confusion is great.

It is desirable to calibrate one's efficiency for detecting the flavor of
neutral $K^*$'s.  Charmed particle decays can be very helpful in this
respect.  Charmed particle decays involving neutral $K^*$'s include
$D^0 \to \overline{K}^{*0} \pi^+ \pi^-$ and $D^+ \to \overline{K}^{*0} \pi^+$.
The
flavor of the decaying $D^0$ can be identified through the chain $D^{*+} \to
\pi^+ D^0$, with emission of a characteristic soft pion.  The $D^+$ decay is
self-tagging.  In a hadron collider, charm signals can be found in channels
such as $\bar B \to D^{(*)} \ell \nu_\ell$, which can be studied using
high-transverse-momentum leptons.

Because $J/\psi K^{*0}$ events are so rare, it may be possible to increase the
data sample by not requiring full reconstruction of the $B$ meson. For
instance, a detached $J/\psi$ is guaranteed to originate in a $b$ quark decay,
and any mode of the type $J/\psi K^{*0} X^0$ or $J/\psi K^{*+} \pi^- X^0$,
where all particles come from the same secondary vertex, reveals the flavor of
the decaying neutral $B$ meson. Neutral pions and photons can be missed, but
one must be more careful about neutral strange particles that escape detection.
If $K/\pi$ separation is available, one can use in addition the mode $J/\psi
K^+ \pi^- X^0$, where the kaon and pion are not necessarily in a $K^{*0}$.   In
order to know whether these semi-inclusive modes are useful, it would help to
know the branching ratios for $B^0 \to J/\psi K^{(r)}$, where $K^{(r)}$ denotes
the higher kaon resonances.  Such information can be provided by present
detectors such as CLEO.
\newpage

\centerline{\bf VII.  CONCLUSIONS}
\bigskip

The correlations of charged pions with neutral $B$ mesons is a potential
source of information on the flavor of the produced heavy meson, which in turn
can be useful in searching for CP-violating asymmetries.  We have shown that
correlations with {\it charged} $B$ mesons can provide a helpful calibration
of this method, but only if potential differences between the two cases are
taken into account.  The fragmentation of a $b \bar b$ pair produced in
isolation would not be expected to lead to such differences, as a result
of symmetry under the isospin reflection $I_3 \to -I_3$.

A genuine physics difference between the behavior of charged and neutral $B$'s
can arise if the fragmentation of a $b$ quark involves other quarks ``picked
up'' from the producing system.   This effect might be sensitive to the
transverse momentum and pseudorapidity of the $B$.  For instance, it could well
be an important feature of production at large longitudinal and small
transverse momenta, but might be less important for central collisions with
high transverse momenta.

Several other effects can give rise to apparent differences between charged and
neutral $B$'s which are in fact spurious.  These include the misidentification
of kaons produced nearby in phase space, the confusion of $B$ decay products
with pions from the primary production vertex, and the interchange of pions and
kaons in forming neutral $K^*$'s.  We have suggested several ways in which one
might test for such effects.
\bigskip

\centerline{\bf ACKNOWLEDGMENTS}
\bigskip

We thank F. DeJongh, P. Derwent,
J. Incandela, E. Kajfasz, M. Shochet, F. D. Snider, S.
Stone, and D. Stuart for discussions. J. R. wishes to thank the Fermilab Theory
Group for their warm hospitality.  This work was supported in part by the
United States Department of Energy under Contract No. DE FG02 90ER40560.

\def \prd#1#2#3{{\it Phys. Rev.} D {\bf#1}, #2 (#3)}
\def \prl#1#2#3{{\it Phys. Rev. Lett.} {\bf#1}, #2 (#3)}
\def \stone{{\it B Decays}, edited by S. Stone (World Scientific, Singapore,
1994)}


\begin{thebibliography}{9}
\bibitem{EX} An exotic mesonic channel is one which cannot be reached from
a quark-antiquark state.
\end{thebibliography}
\end{document}